# Anisotropic dielectric and ferroelectric response of multiferroic LiCu$_2$O$_2$ in magnetic field

Li Zhao, Kuo-Wei Yeh, Sistla Muralidhara Rao, Tzu-Wen Huang, Phillip Wu, Wei-Hsiang Chao, Chung-Ting Ke, Cheng-En Wu, and Maw-Kuen Wu[a]

*Institute of Physics, Academia Sinica, Taipei 11529, Taiwan*

**Abstract** – LiCu$_2$O$_2$ is the first multiferroic cuprate to be reported and its ferroelectricity is induced by complex magnetic ordering in ground state, which is still in controversy today. Herein, we have grown nearly untwinned LiCu$_2$O$_2$ single crystals of high quality and systematically investigated their dielectric and ferroelectric behaviours in external magnetic fields. The highly anisotropic response observed in different magnetic fields apparently contradicts the prevalent *bc*- or *ab*- plane cycloidal spin model. Our observations give strong evidence supporting a new helimagnetic picture in which the normal of the spin helix plane is along the diagonal of CuO$_4$ squares which form the quasi-1D spin chains by edge-sharing. Further analysis suggests that the spin helix in the ground state is elliptical and in the intermediate state the present *c*-axis collinear SDW model is applicable with some appropriate modifications. In addition, our studies show that the dielectric and ferroelectric measurements could be used as probes for the characterization of the complex spin structures in multiferroic materials due to the close tie between their magnetic and electric orderings.

Recent discoveries of the strong coupling between magnetism and ferroelectricity (FE) in some manganites, stimulated the revival of the research on multiferroics, thus bring new opportunities for breakthroughs in both fundamental physics and potential applications [1-3]. The FE found in these materials is of magnetic origin. Most of them are frustrated magnets with complex non-collinear spin structures [2]. One of the commonly accepted microscopic mechanisms comes down to the inverse Dzyaloshinskii-Moriya (DM) interaction, i.e. an antisymmetric relativistic correction to the superexchange coupling [4]. It can also be expressed in an equivalent spin current picture proposed by Katsura, Nagaosa and Balatsky (the so-called KNB model) [5]. The microscopic polarization induced by neighbouring spins is formulated as $\mathbf{P}_{ij}=A\hat{e}_{ij}\times(\mathbf{S}_i\times\mathbf{S}_j)$, where the coupling coefficient $A$ is determined by the spin-orbit coupling and exchange interactions. $\hat{e}_{ij}$ is the unit vector connecting sites *i* and *j*. This expression is also consistent with the phenomenological theory based on symmetry consideration [6], and has worked well in the case of helimagnetic TbMnO$_3$ and other multiferroics [7]. But at present the accurate prediction of the magnitude of the polarization is difficult because much more complicated factors must be considered in real systems.

LiCu$_2$O$_2$ is the first multiferroic cuprate to be reported and is also a prototype of the "1D spiral magnetic ferroelectrics" [8]. As shown in Fig. 1, it has an orthorhombic crystal structure with equal number of Cu$^{2+}$(**s**=1/2) and nonmagnetic Cu$^+$ ions. Cu$^{2+}$ ions locate in the center of CuO$_4$ squares, which form spin chains along the b-axis by edge-sharing. These chains are well separated by coplanar chains of Li$^+$ ions and intercalation layers of Cu$^+$ ions. Along the CuO$_4$ chains, the Cu$^{2+}$-O$^{2-}$- Cu$^{2+}$ bond angle is nearly 90°.

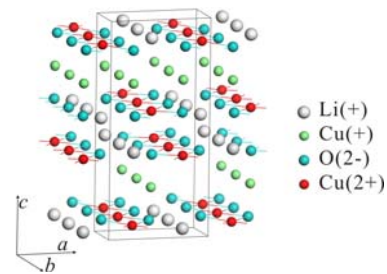

Fig. 1: (Colour on-line) Crystal structure of LiCu$_2$O$_2$.

According to the Kanamori-Goodenough rule, the exchange interaction between the nearest-neighbor Cu$^{2+}$ spins is weak ferromagnetic(FM) (J$_1$<0) in nature while the next-nearest-

---

neighbor interaction is antiferromagnetic(AFM) ($J_2>0$). Such competing interactions can lead to an incommensurate helimagnetic state in classical magnets. However, in quantum spin systems, strong frustration can induce a spin-liquid state with only short-range spin correlations [9, 10]. In $LiCu_2O_2$, $|J_2/J_1|$ is just a little larger than the critical value, and the classical helical magnetism survives out of the quantum liquid case in the end as observed in neutron experiments [11]. There have been many experiments on $LiCu_2O_2$ [8, 11-16]. Two successive magnetic transitions occur at low temperature ($T_{N1}\sim 25K$ and $T_{N2}\sim 23K$ respectively). The intermediate state ($T_{N2}<T<T_{N1}$) is generally considered to be an antiferromagnetic spin-density-wave (SDW) state with *c*-axis collinearity. When $T<T_{N2}$, $LiCu_2O_2$ goes into a non-collinear helimagnetic ground state [11], and concomitantly the *c*-axis spontaneous electric polarization ($P_c$) arises [8].

However, the magnetic structure of $LiCu_2O_2$ in ground state, is still under debate today. Based on their neutron diffraction experiments, Masuda *et al* [11] first proposed a magnetic structure consisting of a transverse-spiral (cycloidal) modulation along the spin chain direction (*b*-axis) with $Cu^{2+}$ spin spirals in the *ab*-plane. However, the existence of $P_c$ contradicts with either the KNB model or the inverse DM mechanism, since the polarization generated by the cycloidal order should be in the spin spiral plane and perpendicular to the chain theoretically. Later in 2007, Park *et al* [8] proposed another spin picture in which the spins spiral in the *bc*-plane rather than the *ab*-plane. The subsequent polarized neutron scattering experiments partially supported this model, but there was considerable discrepancy from the quantitative calculation of the intensity of neutron reflections based on this simple *bc*-spiral picture. The discrepancy was just attributed to the quantum fluctuation in $LiCu_2O_2$ by Seki *et al* [13]. Recently, based on their own NMR and neutron diffraction data, Kobayashi *et al* [15, 16] proposed a new possible spin model consisting of a ellipsoidal spin helix with the in-*ab*-plane helical axis tilted by about 45° from the *a* or *b* axis. There is, however, no further experimental support for this picture so far. The conventional methods for magnetic structural characterization seem to be incapable of determining completely the complex magnetic ground state of $LiCu_2O_2$.

It is also notable that most of the $LiCu_2O_2$ samples used in previous experiments were grown by the flux method [17], which were naturally *ab*-plane twinned since *a* is close to *2b* in $LiCu_2O_2$. The twinning problem adds much more complexity to data analysis due to the inhomogeneous magnetic states at domain walls, which can also induce a local polarization [2, 6]. As a result, untwinned crystals are urgently needed to clarify this issue. Now the advances in the crystal growth techniques have enabled us to avoid this twinning problem in $LiCu_2O_2$ greatly [18]. Herein, we report the dielectric and ferroelectric properties of our nearly untwinned $LiCu_2O_2$ single crystals in different magnetic fields. The concrete results as presented here enable us to reconsider all the possible spin models of $LiCu_2O_2$.

Crystals used in the present experiments were grown in a optical floating zone furnace with two 1.5 kW halogen lamps (model 15HD, NEC Nichiden Machinary) [18].. The growth parameters have been optimized to reduce the twinning in $LiCu_2O_2$ crystals greatly and the corresponding details will be reported elsewhere. The shinny plate-like crystals with highly *c*-axis orientation could be easily cleaved from the rods. To characterize the twinning structure in the *ab*-plane of our crystals quantitatively, we performed phi-scan measurements of the (101) Bragg reflection on a four-circle diffractometer using the synchrotron beams in National Synchrotron Radiation Research Center (NSRRC).

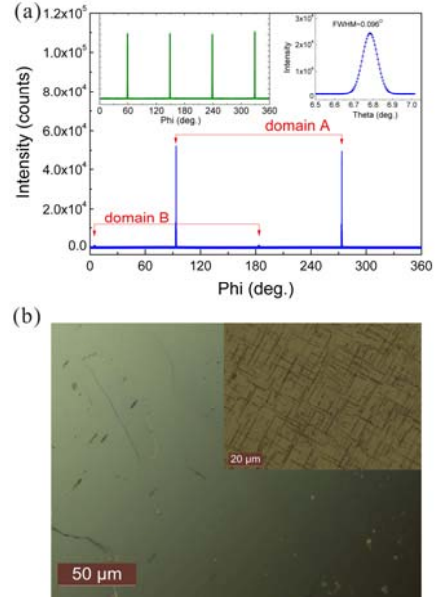

Fig. 2: (Colour on-line) (a) Phi-scan of a typical optimal $LiCu_2O_2$ single crystal measured in NSRRC. The upper-right inset is the rocking curve around the (101) Bragg reflection peak of the domain A with the FWHM of 0.096°. For comparison, the phi-scan of another flux-grown crystal with a severe twinned structure is shown in the upper-left inset.   (b) The polarized optical image of the surface of the optimal untwinned crystal taken with DIC. The image taken from a heavily twinned sample is shown in the inset for comparison.

As for our optimal sample (shown in Fig. 2(a)), the main two-fold symmetry indicates that only one single domain (named "domain A") dominates in this sample. There also exists a trace amount of another set of diffraction peaks, which comes from a minor domain ("domain B") that is *ab*-in-plane orthogonal to the domain A. According to our quantitative estimate, the domain B is much less than 5% in our optimal samples. The FWHM of the rocking curves (shown in the upper-right inset) around the (101) Bragg reflection from domain A is less than 0.1°, indicating that the in-plane mosaic structure is strongly suppressed. The corresponding crystal surface images are taken in the polarized light using a metalloscope (model DM2500, Leica Microsystems Ltd.) with a differential interference contrast (DIC). The optimal sample shows a large smooth untwinned area with scarcely visible grain boundaries.

For comparison, the heavily twinned flux-grown crystals were also examined. Although these samples exhibit flat shinny surfaces, the corresponding phi-scan results (shown in the upper-left inset in Fig. 2(a)) reveal the coexistence of two kinds of nearly-equal domains, which are orthogonal to each other in the *ab*-plane. The polarized optical measurement is shown in the inset of Fig. 2(b). The mutually orthogonal boundaries, which divide the different domains, are along the diagonal of $CuO_4$ squares according to our X-ray diffraction studies, i.e. 45° away from the crystallographic *a*- or *b*- axis (very close to [1, 2, 0] and [1, $\bar{2}$, 0] directions).

It should be noted that the beam size we used in NSRRC is $2\times 0.2mm^2$ in both horizontal and vertical directions and the x-ray energy is 10keV. According to the absorption coefficient of $LiCu_2O_2$ at this energy, the penetration depth of the incident beam should reach several microns. Under this condition, our characterizing technique is obviously not a "local" or "superficial" probe. Thus, these results show that our optimal samples are nearly un-twinned single crystals of high quality.

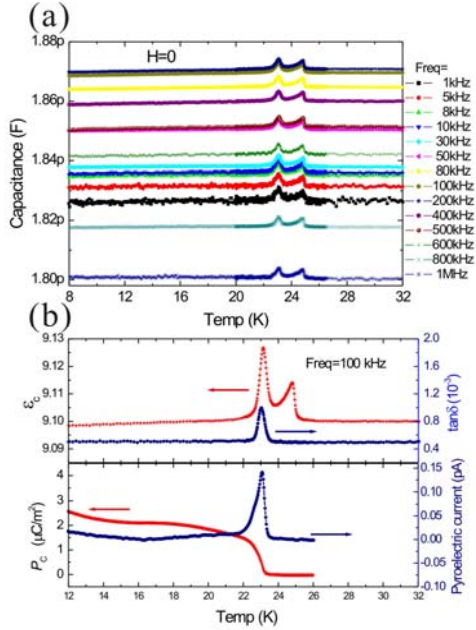

**Fig. 3:** (Colour on-line) (a) Raw capacitance data vs temperature of a typical $LiCu_2O_2$ sample measured at different frequencies in H=0. (b) Upper: $\varepsilon_c(T)$ (red) and corresponding tan loss(blue) measured at a frequency of 100kHz in H=0. Bottom: pyroelectric current (blue) measured at a warming rate of 3K/min in H=0. The poling electric field is about 400kV/m. The corresponding polarization (red) is obtained by integrating the pyroelectric current from above $T_{N1}$.

To perform the *c*-axis dielectric constant and spontaneous polarization measurements, silver paint was applied to end surfaces of the optimal sample as electrodes to form a parallel-plate-like capacitor, whose capacitance is proportional to the *c*-axis dielectric constant ($\varepsilon_c$). $P_c$ is obtained by integrating pyroelectric current. We also measured other optimal samples taken from different batches, and no apparent difference was observed.

Fig. 3(a) shows the raw capacitance data obtained from a typical optimal sample in H=0 measured at a very slow warming sweeping rate (0.1-0.5 K/min). Two sharp peaks are observed for all the testing frequencies(from 1 kHz to 1 MHz), which represent to the two successive magnetic transitions in $LiCu_2O_2$ at $T_{N1}$ (~24.8K) and $T_{N2}$ (~23K) observed in the corresponding magnetization measurements (not shown here). For convenience, we just adopt the data at 100 kHz in the following discussions. As shown in Fig. 3(b), the peak of dielectric loss (tanδ), occurs only around $T_{N2}$, suggesting a proper FE transition. $P_c$ also arises only below $T_{N2}$, which is consistent with the previous reports [8, 12, 13, 15].

As a first step, we applied magnetic field along the *a*-axis (denoted as $H_a$), i.e., perpendicular to the spin chains in $LiCu_2O_2$. As seen in Fig. 4(a), the field-induced change of $\varepsilon_c$ is negligible till $H_a$=9T when $T>T_{N1}$, indicating the absence of magnetoelectric coupling above the magnetic ordering temperature. At $T<T_{N2}$, the entire dielectric background is suppressed with increasing $H_a$. A weak hump structure is observed when $H_a \geq 4T$, which shifts to a higher temperature as $H_a$ increases. Finally, the hump merges into the dielectric peak around $T_{N2}$ at $H_a$ =7T.

L. Zhao  et al.

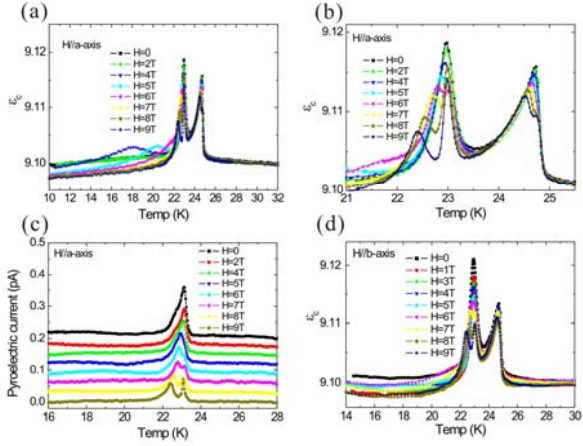

Fig. 4: (Colour on-line) (a) $\varepsilon_c$(T) measured with different $H_a$ (0-9T). Details around the two dielectric anomalies are re-plotted in (b) for clarity. (c) Pyroelectric currents measured at a warming rate of 3K/min in different $H_a$ =0-9T. The curves have been shifted vertically for clarity. (d) $\varepsilon_c$(T) measured with $H_b$(0-9T).

The most striking feature is the field-induced splitting of the two dielectric anomaly peaks, which are re-plotted in Fig. 4(b) for clarity. The splitting develops with increasing field. As $H_a$ increases, around $T_{N2}$, , one of the two sub-peaks moves further to a lower temperature while the other shifts slightly towards higher temperature. The field-induced splitting also exists around $T_{N1}$, but it is much weaker. Although this splitting is quite small (less than 1K), they were confirmed many times in our accurate measurements on several samples taken from different batches. The splitting of the ferroelectric transition at $T_{N2}$ is also verified in our pyroelectric measurements. The measured pyroelectric current at a fixed warming rate is proportional to the temperature derivative of $P_c$ (d$P_c$/dT), and clearly reveals the fine changes of $P_c$(T) induced by external magnetic field. As shown in Fig. 4(c), the splitting at $T_{N2}$ is also observed to develop with increasing $H_a$, consistent with the corresponding dielectric measurements. To our knowledge, it is the first observation of the field-induced splitting of the FE transitions in the multiferroic materials.

Results obtained with H applied along b-axis (denoted as $H_b$, parallel to the spin chains) are shown in Fig. 4(d). The splitting of two dielectric peaks still exists and their corresponding evolution with increasing $H_b$ is quite similar as that in $H_a$. The results of pyroelectric measurements in $H_b$ are almost the same as those in $H_a$ (not shown here). The only noticeable difference between the $H_a$ and $H_b$ cases is the absence of the hump structure in $H_b$ up to 9T at $T<T_{N2}$. This suggests that the hump arises from the enhanced inter-chain coupling caused by applying a transverse field to the quasi-1D spin chains along the b-axis.

In the quasi-1D ab- or bc-in-plane cycloidal spin picture, the minor difference observed in $H_a$ and $H_b$ seems rather puzzling and will be discussed later. Stimulated by a new model proposed by Kobayashi et al [16] (shown in Fig. 6(c)), we further applied magnetic field along the c-axis and the diagonal of the CuO$_4$ squares (the O-Cu-O bond direction), denoted as $H_c$ and $H_{dia}$ respectively. The corresponding results are shown in Fig. 5. The diagonal direction is very close to the [1, 2, 0] or [1, $\bar{2}$, 0] orientation since $a \sim 2b$. It must be emphasized that the fields along [1, 2, 0] and [1, $\bar{2}$, 0] are found just equivalent in our experiments so far as the sample is first cooled in $H_{dia}$ to low temperature to lift the degeneracy. The degeneracy problem in LiCu$_2$O$_2$ will be discussed elsewhere.

As shown in Fig. 5(a), up to the highest field $H_c$ =9T, no splitting effect is observed in $\varepsilon_c$(T) around $T_{N1}$ and $T_{N2}$. In the corresponding measurements of $P_c$, there is neither splitting nor broadening of the pyroelectric current peak as shown in Fig. 5(b). When $H_c$ increases, the dielectric peak at $T_{N1}$ gradually shifts to lower temperatures with decreasing magnitude, indicating the suppressed c-axis AFM transition from the high-temperature paramagnetic state to the intermediate state. Conversely, around $T_{N2}$, the peak of $\varepsilon_c$(T) moves to higher temperature and its amplitude rises sharply as $H_c$ increases, indicating a field-enhanced FE transition. Correspondingly, $P_c$ increases greatly with growing $H_c$. In Fig. 5(b), the $P_c$ at 14K in $H_c$ =9T is 6.4μC/m$^2$, which is nearly 3 times that in zero field (2.2μC/m$^2$). The enhancement of $P_c$ by $H_c$ was also reported by Park et al [8], but much weaker than our results (the $P_c$ in their sample increases about 50% as $H_c$ increases from 0 to 9T, see Fig. 4(c) in Ref. 8), possibly due to the severely twinned structures in flux-grown crystals which obscure the intrinsic properties of LiCu$_2$O$_2$ in some way. Although $H_c$ is also transverse to the spin chains as $H_a$, the dielectric hump structure is absent till 9T as $T<T_{N2}$, indicating the inter-chain coupling interactions are weaker along the c axis than in the ab-plane.

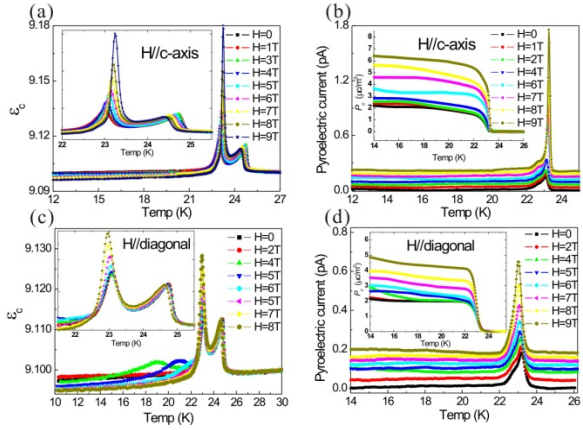

Fig. 5: (Colour on-line) (a) and (c): $\varepsilon_c$(T) measured at different $H_c$ and $H_{dia}$ from 0 to 9T respectively. The details around the two successive transitions are re-plotted in the respective insets respectively. (b) and (d) are corresponding pyroelectric currents measured at a fixed warming rate of 3K/min (the curves are shifted vertically for clarity). The calculated polarizations, $P_c$(T), are also shown in the respective insets.

On the other hand, when $H_{dia}$ is applied, there is also no field-induced splitting (shown in Fig. 5(c) and (d)) as in $H_c$. However, considerable difference exists between the two cases. $H_{dia}$ has almost negligible influence on the first transition at $T_{N1}$, consistent with c-axis SDW picture of the intermediate state. In this picture, the $Cu^{2+}$ spins order antiferromagnetically along the *c*-axis and thus orthogonal to the *ab*-in-plane $H_{dia}$. The magnitudes of dielectric anomaly and pyroelectric peak at $T_{N2}$ increase with growing $H_{dia}$, suggesting the field-enhanced FE transition. However, the transition temperature $T_{N2}$ (labeled by the peak in $\varepsilon_c$(T)) is shifted slightly to a lower temperature by $H_{dia}$, an inverse version of the $H_c$ case (compare the insets in Fig. 5(a) and (c)). Furthermore, the enhancement of ferroelectricity by $H_{dia}$ is not so remarkable as that by $H_c$. $P_c$ at 14K in $H_{dia}$ =9T is 4.85μC/m$^2$, about 2 times that with H=0. When T < $T_{N2}$, the hump structure appears and its evolution with $H_{dia}$ resembles that of the $H_a$ case, further confirming the enhancing effect of the in-plane transverse field component on the inter-chain interactions.

We summarize all our results on the field-induced shift of the two successive transition temperatures ($\Delta T_{Ni}$(H)= $T_{Ni}$(H) - $T_{Ni}$(H=0), i=1, 2) in Fig. 6 (a) and (b) respectively. The $T_{N1}$ and $T_{N2}$ are defined as the maxima of the two dielectric anomaly peaks (around 24.8 and 23K, respectively). The largest splitting is only 0.6K (as $\Delta T_{N2}$ in $H_c$ =9T). Our highly accurate measurements guarantee the probing of the very small shift of the two transitions induced by external magnetic fields with highly reproducibility.

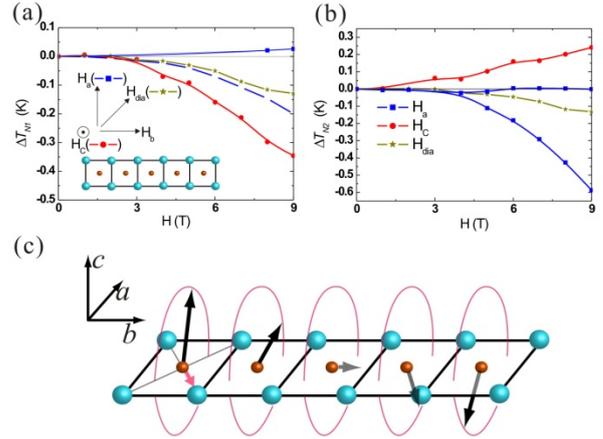

Fig. 6: (Colour on-line) (a) and (b): plots of the field-dependent temperature shift of the two successive magnetic transitions at $T_{N1}$ and $T_{N2}$ respectively. The different field configurations are also illustrated in the inset of (a). (c): Schematic view of the spin configuration of LiCu$_2$O$_2$ in the ground state. The blue and red balls represent $O^{2-}$ and $Cu^{2+}$ ions respectively. The pink arrow represents the normal of the spin spiral plane and black arrows stand for the spins of s=1/2 $Cu^{2+}$ ions.

In our magnetization measurements at low temperature (not shown here), linear M-H curves are observed up to 7T (the limit of our Quantum Design SQUID VSM magnetometer) regardless of the H direction, which coincides with the recent results by Kobayashi et al [15]. This suggests that the basic magnetic structure in LiCu$_2$O$_2$ holds robust even in a very strong magnetic field and no metamagnetic transition such as spin flop or flip occurs. The splitting of the transitions at $T_{N1}$ and $T_{N2}$ in $H_a$ or $H_b$ cannot be ascribed to the emerging of new magnetic phases. In addition, the M-T curves show negligible difference with H along *a*- and *b*- axis, consistent with recent report on untwinned crystals [18].

The present results can not be interpreted in the *ab*- or *bc*- plane spin cycloidal model. In these models the spin configuration is symmetric to the crystallographic *a*- or *b*- axis. Intuitively, if these models stand, the splitting should not occur with H along the symmetric axes (*a* or *b*), but may appear only when H is along the diagonal direction. Since $H_{dia}$ can be considered as two decomposed components, along and transverse to the spin chains, $H_{dia}$ would have different (even adverse) effects that coexist, and may in turn lead to the splitting. Furthermore, if either *ab*- or *bc*- spin spiral model holds, there should be considerable difference between the $H_a$ and $H_b$ cases, which is not observed in the present experiments by us and other research group [18].

All our results, especially the splitting of the transitions in $H_a$ and $H_b$, but not in $H_c$ and $H_{dia}$, suggest that in the ground state the spin spiral planes contain neither crystallographic *a* nor *b* axis. Instead, they contain the mutually orthogonal *c*-axis and the diagonal of the $CuO_4$ square. Although more work (especially theoretical explanation) is needed, based on the symmetry consideration we can still reach the conclusion that the exotic magnetic structure of $LiCu_2O_2$ at low temperature is symmetric with respect to these *c* and diagonal axes as argued above, rather than crystallographic *a*- or *b*- axis of the orthogonal lattice. The present experimental results strongly support the newly-proposed 45°-tilt model by Kobayashi *et al* [16], which is shown in Fig. 6(c) schematically. Although there have been some theoretical work on $LiCu_2O_2$, this new spin model is neglected to some extent (we haven't found any corresponding calculation based on this model). But at present, it may be the only model that is consistent with our results.

In this new picture, *a*- or *b*- axis is no longer symmetric axis of magnetic structure in ground state of $LiCu_2O_2$. As mentioned earlier, when $H_a$ or $H_b$ is applied, external magnetic field can be decomposed into two equal components along the two intrinsic symmetry axes of spin configuration respectively. One component is along the normal to the spin helix plane and the other is in the spin helix plane. Both can cause slight distortions of the ideal spin spiral configurations. Obviously, the effects of these two kinds of distortion are qualitatively different. One is the "cone-like" distortion of spin helixes (from the former out-of-helix-plane component of external magnetic filed as $H_a$ or $H_b$), and the other is due to the inhomogeneity of phase difference between neighboring spins (from the latter in-helix-plane component), as in other helimagnetic systems [19]. Apparently, the former cone-like distortion will reduce the *c*-axis polarization according to the spin current model or inverse DM mechanism, while the effect of the latter distortion is more complicated. The observed splitting can be attributed qualitatively to the coexistence of these two different components.

On the other hand, no splitting effect occurs when the external magnetic field is applied along the intrinsic symmetry axis of the spiral magnetic structure (as in $H_c$ or $H_{dia}$). Only the latter effect exists. Even a slight distortion from the ideal spin spiral configuration can lead to a markedly enhancement of $P_c$, because $P_c$ is the bulk average of $\boldsymbol{P}_{ij}$, which depends greatly on the phase difference between neighboring spins according to the KNB formula. Although $H_c$ and $H_{dia}$ are both in the spiral plane and enhance the multiferroicity greatly, they have quantitatively different influence on FE as discussed earlier. This difference suggests the presence of a noticeable anisotropy, which is also consistent with the ellipticity of the spin helix structure proposed by Kobayashi [16]. Further detailed work is underway.

The corresponding down- or up-shift of the FE transition temperature ($T_{N2}$, marked by the dielectric peaks) is driven by the change of total free energy, including magnetic exchange energy between neighboring spins and their interactions with external field. Further quantitative calculation of the microscopic magnetic energetics is needed and the corresponding calculation requires a detailed structure of the field-distorted spin configurations and the corresponding exchange energies, which can be examined in further high-resolution neutron experiment under magnetic field.

In the intermediate state ($T_{N2}<T<T_{N1}$), the *c*-axis collinear SDW picture was widely accepted, and consistent with the absence of FE and the observed suppression of $T_{N1}$ by $H_c$. However, this picture contradicts the observed splitting of the transition at $T_{N1}$ in $H_a$ or $H_b$. The splitting is absent in $H_c$ or $H_{dia}$, something similar to the multiferroic transition at $T_{N2}$. So the present "pure" collinear picture needs some necessary and appropriate revision. One most possible proposal is that all the spins of $Cu^{2+}$ ions lie in the planes whose normals are the diagonals of $CuO_4$ squares. The spins are preferentially aligned along the positive or minus *c*-axis with slight in-plane angular fluctuation. This quasi-collinear "fluctuating" SDW picture is quite consistent with the present results. The fluctuation planes evolve into the spin-helix planes as the sample is cooled below $T_{N2}$. Further neutron diffraction measurements are needed to confirm our conjecture.

Additionally, the weak hump in the $\varepsilon_c(T)$ at $T< T_{N2}$ is observed only when the inter-chain interactions are enhanced by a strong transverse field component in the *ab*-plane (as in $H_a$ or $H_{dia}$). Besides the hump structure, no remarkable difference is observed between the $H_a$ and $H_b$ cases, indicating that the inter-chain interactions have almost negligible effect on the multiferroicity. The 2D-like characteristics of $LiCu_2O_2$ observed in some previous experiments may come from the mixed response when the external excitations were applied along the *a*- or *b*- axis [14]. Our results show that the strong quasi-one dimensionality existing in the $LiCu_2O_2$, is important for further theoretical modeling.

In summary, the anisotropic dielectric and ferroelectric response is observed in nearly untwinned $LiCu_2O_2$ single crystals in different external magnetic fields, which enables us to deduce its complex magnetic structures. Our results provide strong evidence in support of the new helical model proposed by Kobayashi *et al* [16]. As for the intermediate state ($T_{N2}<T<T_{N1}$), a tentative modification is suggested to the present collinear SDW model to meet our new findings. The huge enhancement of FE from slight distortion of its spin configuration in certain fields ($H_c$ and $H_{dia}$) may provide a new clue to develop materials with stronger multiferroicity. Our studies highlight the importance of dielectric properties in the characterization of complex magnetic structures in multiferroic materials due to the strong tie between their magnetic and electric orderings.

***

We thank Dr M.J Wang, W.L. Lee, C.L. Chen and M.P.L. Wu for their technical support and helpful discussions